# STUDY OF SEY DEGRADATION OF AMORPHOUS CARBON COATINGS*

N. Bundaleski#, S. Candeias, A. Santos, O.M.N.D. Teodoro and A.G. Silva, CEFITEC, Departamento de Física, Faculdade de Ciências e Tecnologia, Universidade Nova de Lisboa, 2829-516 Caparica, Portugal

*Abstract*

Deposition of low secondary electron yield (SEY) carbon coatings by magnetron sputtering onto the inner walls of the accelerator seems to be the most promising solution for suppressing the electron cloud problem. However, these coatings change their electron emission properties during long term exposure to air. The ageing process of carbon coated samples with initial SEY of about 0.9 received from CERN is studied as a function of exposure to different environments. It is shown that samples having the same initial SEY may age with different rates. The SEY increase can be correlated with the surface concentration of oxygen. Annealing of samples in air at 100-200 °C reduces the ageing rate and even recovers previously degraded samples. The result of annealing is reduction of the hydrogen content in the coatings by triggering its surface segregation followed by desorption.

## INTRODUCTION

Electron cloud (e-cloud) represents one of the major problems in achieving high quality beam in modern particle accelerators [1, 2]. Its development can induce emittance growth, bunch-to-bunch coupling and relevant thermal load in case of cryogenic machines. The process starts with electrons initially generated by ionization of the residual gas, by beam losses in the walls and by photoelectrons. When such electrons impinge on the walls of the vacuum pipe, they may be multiplied and interact with the high-energy beam, reducing its quality. The problem can be suppressed by reducing the secondary electron yield (SEY) of the inner walls of the system, defined as the number of emitted electrons per incident electron. For instance, in the case of LHC, if the SEY is below the threshold of 1.3 the electron cloud can be completely suppressed [3].

Among several approaches to solve this problem, amorphous carbon (a-carbon) coatings deposited by magnetron sputtering appears to be the most promising solution [4, 5]. They have excellent ability in maintaining the SEY below unity, even after a few hour exposure to air, without any bake-out. While initial tests on e-cloud monitors in the SPS (Super Proton Synchrotron) have demonstrated that such coatings can suppress e-cloud [4, 6] it appeared that SEY increase takes place due to long term air exposure.

In our previous work we showed that these samples do not age only in air but also under high vacuum (HV) conditions [7]. Furthermore, ageing in HV was actually far more pronounced. The most striking result was observation of ageing of samples exposed to low $10^{-9}$ mbar obtained by oil diffusion pumps. Careful analysis of this problem revealed that HV and even UHV ageing can take place depending on the characteristics of the pumping system – oil of pumps appears to be a contaminant responsible for ageing in vacuum.

Having in mind that the equipment in CERN Technology department used for testing the samples does not differ much from the system in which our experiments were performed (HV in CERN is achieved with a turbo pump with rotary oil backing pump working without the oil filter [8]), there were still some contradictions between the results obtained in the two groups. In this paper we resolve this paradox, showing that samples having same SEY may indeed have different ageing properties. The latter is probably related to some contamination that does not change SEY directly, but contributes to faster ageing. The nature of ageing process in HV is further studied in this work, as well. In particular, we demonstrate novel and surprising effects of ageing suppression and SEY recovery due to sample annealing in air at temperatures up to 200 °C. Explaining how annealing of a-carbon coatings influences ageing could also help in understanding the ageing in air and HV. All this can provide better specification of vacuum requirements for preserving long term stability of these coatings during the accelerator exploitation.

## EXPERIMENTAL

The amorphous carbon coatings were deposited on technical stainless steel via magnetron sputtering using Ne as the discharge gas. The samples arrived from CERN Technology Department– Vacuum Surfaces and Coatings group in a UHV container filled with 500 mbar of pure nitrogen. They are in the form of thin 100 mm long and 10 mm wide bars. The samples used in experiments were made by cutting the bars to a typical size of 7x10 $mm^2$. After any opening, the chamber would be refilled with nitrogen gas. This way of conditioning the surface suppresses their ageing [8], which we confirmed during a period surpassing one year.

X-ray Photoelectron Spectroscopy (XPS) has been performed in a multipurpose surface analysis system [9] with a base pressure in the low $10^{-10}$ mbar range. A homemade apparatus for SEY measurements was mounted on a load-lock system used to introduce samples

*Work supported by Fundação para a Ciência e Tecnologia do Ministério da Ciência, Tecnologia e Ensino Superior (FCT/MCTES), through the project CERN/FP/109313/2009

# n.bundaleski@fct.unl.pt

into the analysis chamber without breaking vacuum, providing base pressure in the low $10^{-8}$ mbar region. The same chamber was used for HV exposure of a-carbon coatings with a typical pressure in $10^{-7}$ mbar region. The design of the SEY system, which measures total secondary electron yield, is similar to the one described in [10]. Depth profiling of the surface composition has been performed using Time of Flight Secondary Ion Mass Spectroscopy (TOF-SIMS) using upgraded VG Ionex IX23LS set-up based on the Poschenrieder design, with a gallium ion gun for the analysis and oxygen ion gun for the sputtering. During the annealing, which was performed in a standard furnace, the samples were wrapped in aluminium foil to prevent their contamination.

## RESULTS AND DISCUSSION

In our previous work we demonstrated that the samples were heavily contaminated by the oil vapour from rotary pumps [7]. Therefore, the introduction chamber in which the SEY measurements and sample exposure to HV are performed was thoroughly washed, dried by long term annealing and mounted back to the main system. The base pressure in the chamber has been reduced by almost two orders of magnitude with respect to previous experiments. Nevertheless, initial tests showed that ageing of samples exposed to this environment was still noticeable although the rate of SEY increase was significantly reduced. However, further experiments revealed that some of the samples actually do not age. As an illustration, we present in Figure 1 results of SEY measurements of two samples before and after exposure to high vacuum at room temperature. The two samples have practically the same initial SEY. After the exposure to the same HV environment SEY of one of them increased to 1.3 after 5 days, while that of the other sample practically did not change after 20 days of exposure. It should be emphasized that the result of the second sample is fully consistent with the general findings of the CERN group. Obviously, low SEY does not guarantee that the sample is in perfect condition. We do not understand yet the reason why two samples produced in the same deposition experiment have different ageing properties. Initially, it was assumed that the ageing behaviour depends on the exact position of samples inside the container: the samples that do not age are placed on the bottom of the container while those that age faster are much closer to the top. This would have implied, as it was intitially observed, that the samples closer to the top (but also to the valve used for pumping and introduction of nitrogen) are more exposed to eventual contaminants. However, further test showed that this assumption has to be overruled.

In the following, we shall mainly focus on the samples that age i.e. on those positioned closer to the top of the container. In one of the experiments, two samples were cut from the same bar and exposed to HV. Only one of them was previously annealed in air to 100 °C for 20 h. The results of SEY measurements before and after HV exposure are presented in Figure 2, showing that previously annealed sample ages much slower. Therefore, annealing modified the a-carbon thin film in such way that SEY increase was suppressed. As for the samples that do not show any ageing in both, HV and air, annealing up to 200 °C for 20 h has practically no influence on their SEY.

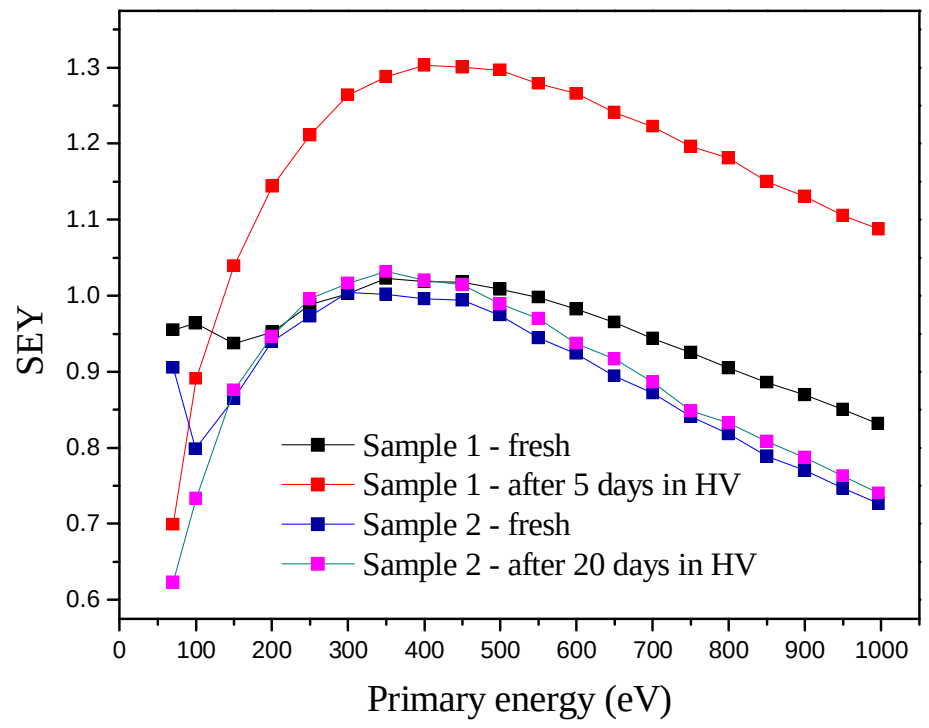

Figure 1: SEY of two a-carbon samples before and after HV exposure. Sample 1 shows ageing after 5 days of exposure while sample 2 does not age even after 20 days of exposure.

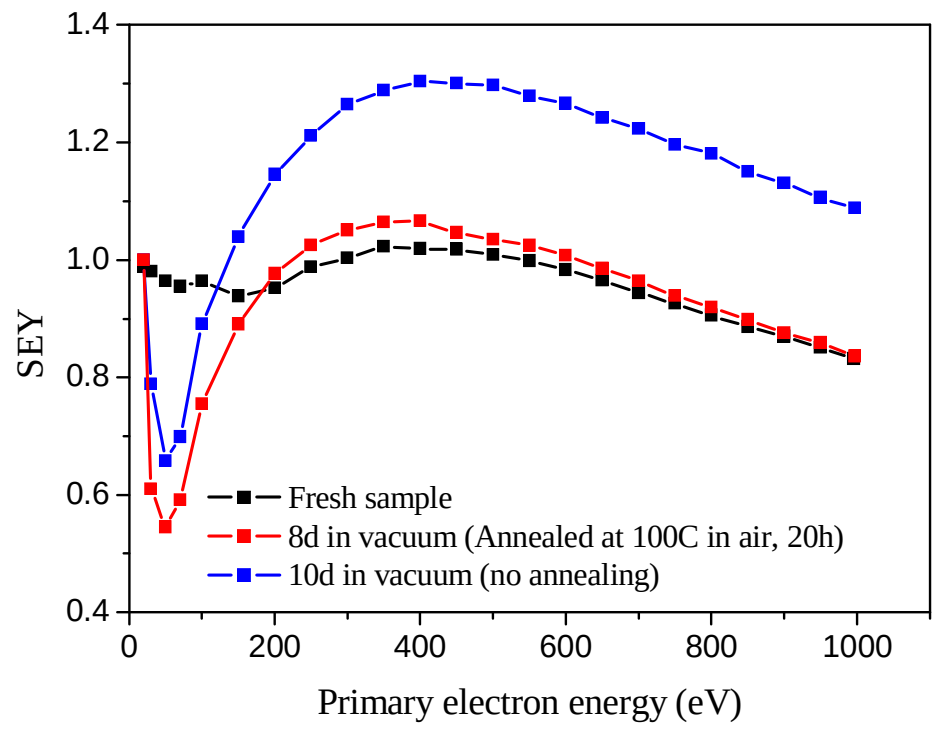

Figure 2: SEY of two samples cut from the same bar, one of which was annealed: the annealed sample ages much slower.

Besides the suppression of ageing, annealing in air can be used to recover SEY of already aged samples. As an illustration, we present in Figure 3 SEY measurement of a sample exposed to air for 14 months before any annealing, after annealing to 120 °C for 3 days, and after annealing to 200 °C for 17 h. As a result of annealing at the higher temperature, maximum SEY was reduced from about 1.9 to less then 1.3.

The results of studying the influence of annealing on ageing suppression and SEY recovery are systematized in Table 1. It can be seen that annealing suppresses ageing in air as well. The annealing temperature necessary for SEY recovery is generally higher than the one needed to suppress ageing, although it also depends on how heavy is the contamination i.e. how high is the initial SEY (see below). We should also emphasize that the samples that did not show ageing in HV, practically did not age after air exposure for 2 months as well.

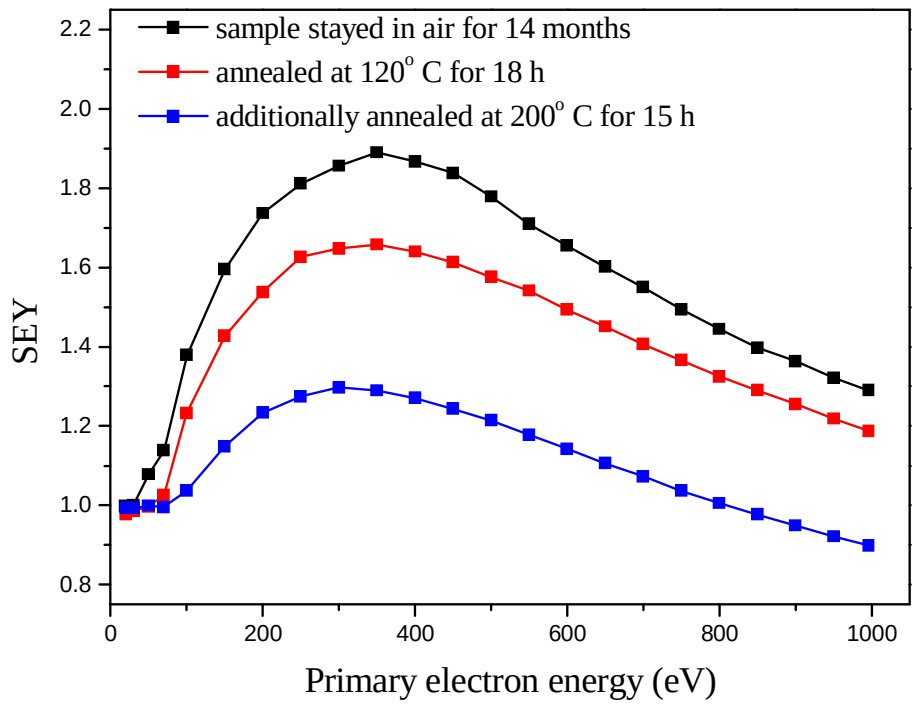

Figure 3: Recovery of secondary electron emission properties by sample annealing to different temperatures.

Table 1 Influence of annealing on ageing rate and SEY recovery: $t_E$ – the exposure time; $t_A$ – the annealing time $T_A$ – the annealing temperature; $\Delta\delta/\delta$ – relative change of the SEY maximum.

| **Influence of annealing on ageing** | | | |
|---|---|---|---|
| **Environment** | **$t_E$** | **$T_A$ (°C), $t_A$ = 20 h** | **$\Delta\delta/\delta$** |
| HV | 10 days | - | 27.9% |
| | 8 days | 100 | 1.46% |
| | 20 days | 100 | 5.74% |
| | 20 days | 200 | 3.4 % |
| Air | 1 month | - | 15.7% |
| | 2 months | 100 | 3.34% |
| **SEY recovery due to annealing** | | | |
| **History** | **$t_A$** | **$T_A$(°C)** | **$\Delta\delta/\delta$** |
| Exposed to HV for several days at T = 100 °C | 18 h | 120 | -2,3% |
| | 2.5 days | 120 | -2.7% |
| | 15 h | 200 | -7.2% |
| Kept in air for about 1 year | 18 h | 120 | -12.3% |
| | 15 h | 200 | -21.8% |
| Kept in HV at r.t. for 20 days | 20 h | 100 | -19% |

In order to improve our understanding of the ageing process, surface composition measurements have been performed using XPS, but also sputter depth profiling using TOF-SIMS. According to the XPS results, besides carbon, there is always a small amount of oxygen at the surface. SEY maximum increases with the oxygen content, as shown in Figure 4. Similar result has been already observed by the CERN group [11]. Additionally, we stress that the oxygen content is decreasing after the annealing.

The most intensive peaks in SIMS spectra are those attributed to carbon and hydrogen, as well as numerous peaks typical for hydrocarbons. Additionally, alkali ions were also observed, which is not a surprise having in mind that SIMS is extremely sensitive to these species in the positive ion mode. Unfortunately, since the sample sputtering was performed using oxygen ion beam, we cannot make any conclusion about oxygen depth distribution. In order to suppress the influence of the matrix as well as of other effects, we were analyzing the intensity of the hydrogen peak relative to that of the carbon. Assuming that this ratio is proportional to the hydrogen content, dependence of this magnitude vs. sputtering time roughly corresponds to the depth distribution of hydrogen in the thin film. The result of the dynamical SIMS analysis of three samples cut from the same bar taken from the top of the container is presented in Figure 5. The first one is the fresh sample (black line). The second sample was exposed to HV for 3 days (blue line), while the third one was annealed after the exposure to HV for 3 days (red line). We clearly see that exposing to HV contributed to the increase of the hydrogen content inside the deposit. The most striking result is that of the annealed sample, which has the same hydrogen content in the bulk as the fresh sample while the hydrogen surface concentration is very high. Assuming that the hydrogen content of the third sample before annealing was the same as that of the second sample, annealing triggered surface segregation of hydrogen and most probably its desorption. Consequently, it seems that at least one of the effects of annealing is removal of hydrogen from the deposit.

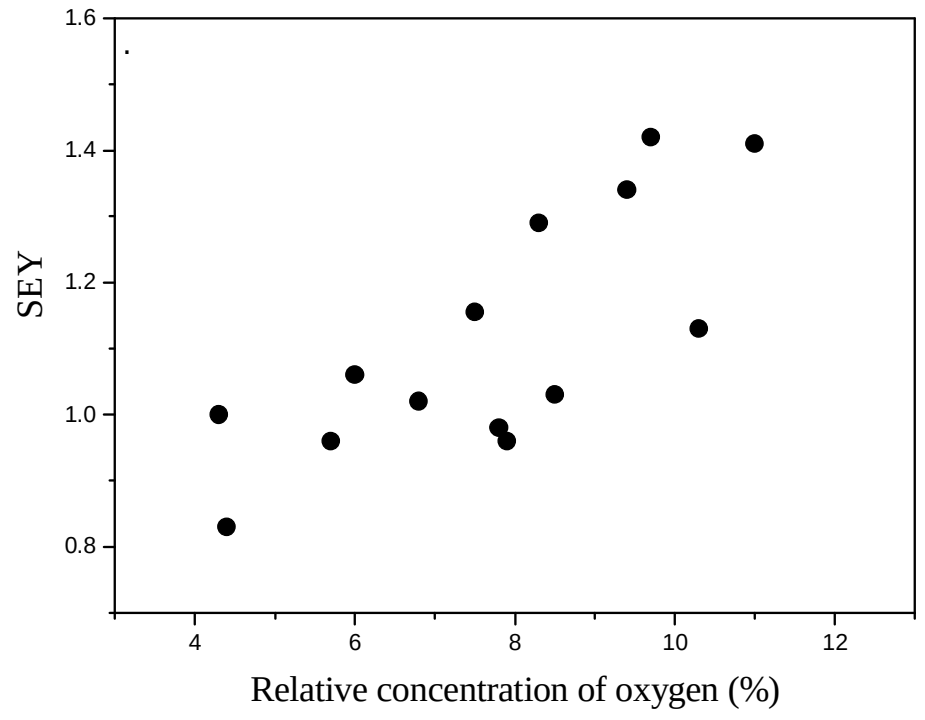

Figure 4: Relative concentration of oxygen vs. SEY

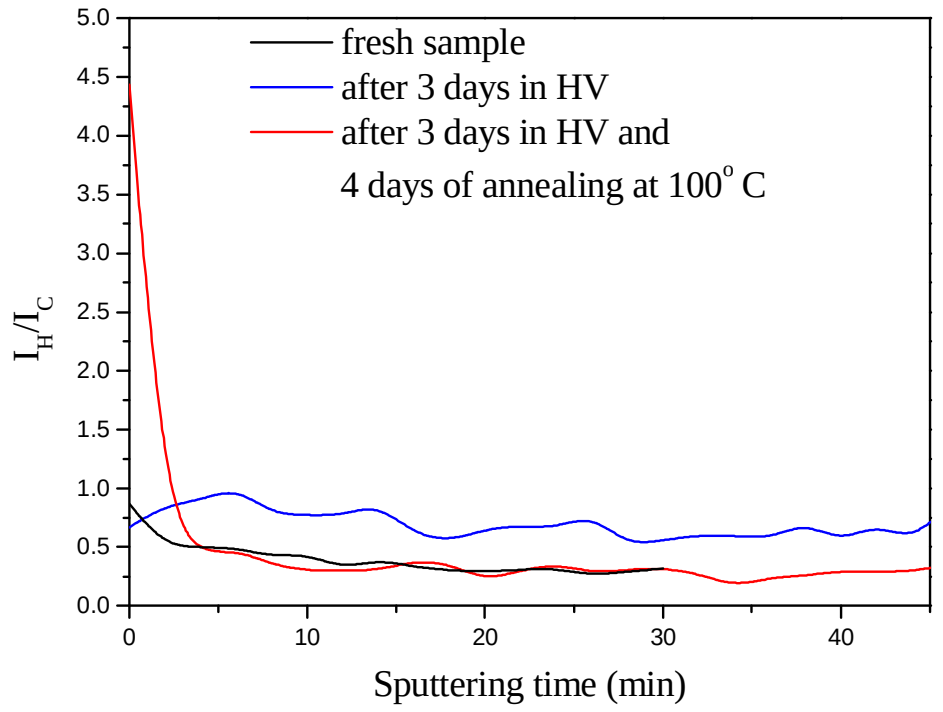

Figure 5: Time dependence of the intensity ratio of hydrogen and carbon peaks in dynamical SIMS measurement.

It is generally accepted that ageing of these samples is caused by changing the hybridization of carbon from $sp^2$ to $sp^3$. Although both, hydrogenation [12] and oxidation [13] of a-carbon coatings can provoke the hybridization change, these chemical processes are highly endothermic and should not take place spontaneously at room temperature. However, one should have in mind that a-carbon is a very complex material with many dangling bonds so that even relative amount of $sp^2$ and $sp^3$ bonds cannot be used to determine its properties [14]. According to the results presented in this work, it appears that SEY degradation takes place only if both, hydrogen and oxygen are present in the coatings. Much more work is necessary to understand well the degradation of a-carbon coatings in terms of secondary electron emission properties.

## CONCLUSIONS

SEY degradation of a-carbon coatings exposed to different environments was studied. It was shown that samples with the same SEY maximum can have different degradation properties. SEY degradation is clearly correlated with the surface concentration of oxygen. Annealing of samples in air at temperatures in the range 100-200 °C decreases degradation rate and even contributes to the sample recovery. The annealing result is surface segregation of hydrogen followed by desorption i.e. the reduction of hydrogen content in the deposit. Apparently, SEY degradation is somehow related to the simultaneous sample contamination by oxygen and hydrogen.

## ACKNOWLEDGMENT

We thank Dr. Alexander Tolstoguzov from Universidade Nova de Lisboa for performing the TOF-SIMS measurements. We acknowledge also funding from the Portuguese research Grant Pest-OE/FIS/UI0068/2011 through FCT-MEC.